# Nonlinear Processes in Multi-Quantum-Well Plasmonic Metasurfaces: Electromagnetic Response, Saturation Effects, Limits and Potentials

J. S. Gomez-Diaz, M. Tymchenko, J. Lee, M. A. Belkin[**], and Andrea Alù[*]

Dept. of Electrical & Computer Engineering, The University of Texas at Austin

[*]alu@mail.utexas.edu, [**]mbelkin@ece.utexas.edu

*Nonlinear metasurfaces based on coupling a locally enhanced plasmonic response to intersubband transitions of n-doped multi-quantum-wells (MQWs) can provide second-order susceptibilities orders of magnitude larger than any other nonlinear flat structure measured so far. Here, we present a comprehensive theory to characterize the electromagnetic response of nonlinear processes occurring in ultrathin MQW-based plasmonic metasurfaces, providing a homogeneous model that takes phase-matching at the unit-cell level and the influence of saturation and losses into account. In addition, the limits imposed by saturation of the MQW transitions on the nonlinear response of these metasurfaces are analytically derived, revealing useful guidelines to design devices with enhanced performance. Our approach is first validated using experimental data and then applied to theoretically investigate novel designs able to achieve significant second-harmonic generation efficiency in the infrared frequency band.*

PACS: 41.20.-q, 42.50.-p, 78.67.De, 73.20.Mf

## I. Introduction

Metamaterial concepts [1] have significantly contributed to the recent research progress in the field of nonlinear optics [2], becoming instrumental to enhance the nonlinearity of optical crystals [3]-[5] by boosting the local density of states and by relaxing the phase-matching constraints in conventional nonlinear processes [6]-[7], thus leading to novel functionalities and applications, such as optical switching and memories [8], super-resolution imaging [9], efficient frequency generation [6],[10]-[11], parametric amplification [12], and bistability [13]. However, the overall response of nonlinear optical metamaterials is still limited and the quest for novel techniques able to provide high conversion efficiencies in nonlinear processes continues [2]. Recently, a novel approach based on properly-designed plasmonic resonances coupled to the intersubband transitions of *n*-doped multi quantum-well (MQW) semiconductor heterostructures [14],[15] has led to ultrathin plasmonic metasurfaces with second-order susceptibilities several orders of magnitude larger than in any other known nonlinear device with subwavelength thickness, constituting the basis for a novel highly-efficient flat platform for nonlinear photonics. This technology has been experimentally demonstrated for second-harmonic generation (SHG) using metasurfaces operating in reflection [14],[15], and it has been theoretically investigated for SHG in transmission [16]. Despite its promising performance, nonlinear MQW-based plasmonic metasurfaces are still in their infancy and they face fundamental and technical challenges that should be properly addressed before this platform may be considered an alternative to well-stablished nonlinear bulk materials [2]. For instance, measured results [14],[15] have shown that the saturation of MQW transitions strongly limits the overall conversion efficiency of these metasurfaces, even for low light intensities. This is consistent with earlier works modelling the nonlinear response of bulk MQW intersubband transitions, which pointed out to strong sensitivity of these effects to saturation and losses [17], [18].

In the present literature, nonlinear metamaterials are usually modelled under the undepleted pump approximation [2] by using coupled-mode theories [19], Green's function approaches [20],[21], Bloch mode expansions [22], or nonlinear scattering theories [23]. In addition, nonlinear processes within MQW-based waveguides have been known for years [24]-[28], and they have been applied to enhance light-matter interactions [29]-[32], to investigate Rabi splitting [33], and to the recent development of quantum-cascade lasers able to generate room-temperature SHG in mid-infrared [34] and terahertz difference-frequency generation (DFG) [35]. In [14], MQWs-based plasmonic SHG metasurfaces operating in reflection and under normal incidence of the input pump were modelled using a homogeneous effective susceptibility tensor based on combining the solution of linear electromagnetic problems through Lorentz reciprocity theorem. The present contribution extends the approach derived in [14] to propose a comprehensive theory able to model the electromagnetic response of nonlinear processes occurring in MQW-based ultrathin plasmonic metasurfaces, rigorously considering the presence of losses and saturation effects, revealing the fundamental limits that saturation imposes to the metasurface optical response, and providing general guidelines to the design of efficient devices with virtually any desired effective nonlinear susceptibility tensor. Specifically, Section II studies the electromagnetic response of periodic nonlinear metasurfaces under normal incidence of the input pump and operating either in reflection or transmission, detailing the derivation of a homogeneous effective nonlinear susceptibility tensor able to characterize a unit cell of the device and forecasting the overall nonlinear response of the entire metasurface. Even though our development focuses on SHG processes, the approach is introduced in a general and systematic way, allowing its direct application to any other nonlinear process, such as differential frequency generation, phase conjugation, sum-frequency generation, etc. Section III derives analytical expressions for the nonlinear susceptibility of ultrathin MQW metasurfaces, rigorously considering the influence of saturation in the nonlinear processes. Section IV describes the fundamental limits, in terms of maximum generated light intensity and conversion efficiency, imposed by saturation to the nonlinear optical response of metasurfaces and provides general design guidelines to minimize its influence. Section V validates the proposed theory against experimental data and then Section VI applies it to design novel SHG metasurfaces with high power conversion efficiency, to investigate how the features of the MQWs affect the performance of the devices, and to predict the maximum performance of the structures. Our results confirm that even sub-optimal resonator designs can achieve large power conversion efficiency, in the order of a percent or more, over deeply subwavelength thicknesses. Further improvements in the plasmonic resonator design are expected to lead to even higher nonlinear conversion efficiencies.

## II. Nonlinear metasurfaces: Effective nonlinear susceptibility tensor and conversion efficiency

Consider a MQW-based nonlinear ultrathin plasmonic metasurface, as schematically illustrated in Fig.1a. In the general case, the metasurface is composed by a periodic arrangement of identical unit cells, each one forming a composite subwavelength resonator – made of dielectrics/metals and MQW layers – able to strongly couple the electromagnetic radiation coming from free-space to the electric field perpendicular ($E_z$) to the MQWs, thus exciting the semiconductor intersubband transitions. It is important to stress that the MQW second-order susceptibility tensor has the form $\bar{\bar{\chi}}^{(2)} = \chi^{(2)}_{zzz}\hat{e}_z\hat{e}_z$ [2],[14],[24], and therefore it only responds to electric fields perpendicular to the quantum-well layers. Consequently, incident light of arbitrarily polarization impinging on bare MQWs will not induce any $E_z$ field, and the semiconductor will not generate any nonlinear response (as experimentally shown in [14]). This highlights the fundamental role of the metasurface subwavelength nanoresonators, which i) provide the required polarization conversion between the incident light and the electric fields induced in the MQWs, and ii) strongly enhance the induced $E_z$ field, thus boosting the intrinsic nonlinear response of MQWs. The advantage of our resonant metasurfaces over traditional nonlinear crystals is that, despite some losses, the metasurfaces can provide considerable frequency conversion in very thin films without the phase-matching constraints of bulk nonlinear crystals. This can be achieved at very low pump intensities, well below materials damage threshold even for continuous-wave operation. Thus, for some experiments, such as autocorrelation measurements or broadband frequency conversion of weak optical fields, our metasurfaces provides advantages over traditional nonlinear crystals.

Nonlinear metasurfaces are modelled here in two separate steps. First, the nonlinear electromagnetic response of the subwavelength unit-cells is retrieved assuming a uniform, arbitrarily-polarized, normally incident field. The proposed approach allows determining an effective transverse susceptibility tensor of the cell and it provides closed-form expressions for the generated intensity and conversion efficiency. This technique can be applied to any nonlinear process/geometry. Second, we consider an actual Gaussian beam impinging on the metasurface and provide the overall nonlinear electromagnetic response of the entire structure.

*IIa. Plane-wave illumination*

The nonlinear response of the unit cells that compose the ultrathin metasurface is evaluated here using the Lorentz reciprocity theorem, assuming a low conversion efficiency of the process, i.e., that the generated intensity is less than 10% of the pump intensity. Saturation effects are inherently taken into account in the formulation but they are introduced later in the text. For the sake of illustration, we focus here on a second-harmonic generation process in a reflective (grounded) unit-cell. The extension of the approach to consider other nonlinear processes and the case of cells operating both in reflection and transmission are also discussed.

Let us consider an electromagnetic wave composed by two harmonics, $\omega_1$ and $\omega_2$, normally impinging on a unit cell of a metasurface (see Fig. 1b). The electric field associated to this wave may be described in free-space as

$$\mathbf{E}_{inc}(\mathbf{r},t) = \mathbf{E}_{inc}^{\omega_1}(\mathbf{r})e^{j\omega_1 t} + \mathbf{E}_{inc}^{\omega_2}(\mathbf{r})e^{j\omega_2 t} + c.c., \tag{1}$$

where 'c.c.' denotes complex conjugate, $\mathbf{r}$ is the vector position, and $\mathbf{E}_{inc}^{\omega_1}(\mathbf{r}) = E_{inc[x]}^{\omega_1}(\mathbf{r})\mathbf{e}_x + E_{inc[y]}^{\omega_1}(\mathbf{r})\mathbf{e}_y$ and $\mathbf{E}_{inc}^{\omega_2}(\mathbf{r}) = E_{inc[x]}^{\omega_2}(\mathbf{r})\mathbf{e}_x + E_{inc[y]}^{\omega_2}(\mathbf{r})\mathbf{e}_y$, where $\mathbf{e}_x$ and $\mathbf{e}_y$ are unit vectors along the directions *x* and *y*, respectively. This incident wave generates an electric field within the MQW region of the cell

$$\mathbf{E}_{MQW}(\mathbf{r},t) = \mathbf{E}^{\omega 1}(\mathbf{r})e^{j\omega_1 t} + \mathbf{E}^{\omega 2}(\mathbf{r})e^{j\omega_2 t} + c.c., \tag{2}$$

that in turn induces a nonlinear polarization in the structure [2]

$$P_i^{\omega_n+\omega_m}(\mathbf{r},t) = \varepsilon_0 \sum_{jk}\sum_{(nm)} \chi_{ijk}^{(2)}(\omega_n+\omega_m,\omega_n,\omega_m;\mathbf{r})E_j^{\omega_n}(\mathbf{r})E_k^{\omega_m}(\mathbf{r})e^{j(\omega_n+\omega_m)t}, \tag{3}$$

where $i,j,k=\{x,y,z\}$, $n,m=\{1,2\}$, the summation over *nm* is performed keeping the sum $\omega_n+\omega_m$ fixed and allowing the factors *n* and *m* to vary, and $\chi_{ijk}^{(2)}(\omega_n+\omega_m,\omega_n,\omega_m;\mathbf{r})$ denotes the components of the nonlinear susceptibility tensor associated to a particular frequency combination [2]. We stress that we adopt the intensity-dependent MQW susceptibility model introduced in [24], as detailed later in Section III. Importantly, since the induced intensity varies across the MQW volume, the susceptibility tensor becomes position dependent. Eq. (3) describes the polarization density induced in the MQW for all nonlinear processes, including SHG, DFG, and sum-frequency generation. In the following, we particularize our analysis to MQW-based metasurfaces designed for SHG. However, the procedure detailed below is general and can easily be applied to any other nonlinear process by simply considering the adequate polarization term.

In the specific case of SHG, the incident field is just composed by a unique harmonic, i.e. $\left|E_{inc}^{\omega_2}\right|=0$. The polarization density induced in the MQW at $2\omega_1$ reads

$$P_i^{2\omega_1}(\mathbf{r},t) = \varepsilon_0 \sum_{jk} \chi_{ijk}^{(2)}(\mathbf{r}) E_j^{\omega_1}(\mathbf{r}) E_k^{\omega_1}(\mathbf{r}) e^{j2\omega_1 t}, \qquad (4)$$

where $\chi_{ijk}^{(2)}(\mathbf{r}) = \chi_{ijk}^{(2)}(2\omega_1,\omega_1,\omega_1;\mathbf{r})$ is the intensity-dependent non-linear susceptibility associated to a SHG process. For the sake of compactness, this last equation is transformed to the frequency domain and simplified taking into account that in MQWs all components of $\bar{\bar{\chi}}^{(2)}(\mathbf{r})$ are zero except the one relating the electric fields perpendicular to the structure [24], $\chi_{zzz}^{(2)}(\mathbf{r})$. Following these steps, Eq. (4) reduces to

$$P_z^{2\omega_1}(\mathbf{r}) = \varepsilon_0 \chi_{zzz}^{(2)}(\mathbf{r}) E_{z[a]}^{\omega_1}(\mathbf{r}) E_{z[b]}^{\omega_1}(\mathbf{r}), \qquad (5)$$

whereas the $x$ and $y$ components of the polarization density are zero and $a,b=\{x,y\}$. In our notation, $E_{z[a]}^{\omega_1}(\mathbf{r})$ refers to the $z$-component of the electric field generated in the position $\mathbf{r}$ within the MQW by an incident electromagnetic wave with angular frequency $\omega_1$ polarized along the direction $a$. The $z$-directed polarization current density generated by the induced polarization is

$$J_z^{2\omega_1}(\mathbf{r}) = j2\omega_1\varepsilon_0\chi_{zzz}^{(2)}(\mathbf{r})E_{z[a]}^{\omega_1}(\mathbf{r})E_{z[b]}^{\omega_1}(\mathbf{r}). \tag{6}$$

In order to relate this current density to the electric field radiated towards free-space at $2\omega_1$, we apply Lorentz reciprocity theorem

$$\int_{V_{UC}}\mathbf{E}_{MQW}^{2\omega_1}(\mathbf{r})\mathbf{J}_{MQW}^{2\omega_1}(\mathbf{r})dV = \int_S \mathbf{E}_{FS}^{2\omega_1}(\mathbf{r})\mathbf{J}_{(2D)}^{2\omega_1}(\mathbf{r})ds, \tag{7}$$

where the subscripts 'UC' refers to the unit-cell volume and the subscript 'S' refers to an infinitesimally thin region, i.e., a sheet with the same area as the unit-cell, in free-space located in the far-field with respect to the MQW. Eq. (7) indicates the relation between the surface current density $\mathbf{J}_{(2D)}^{2\omega_1}$ in the far-field that generates an electric field $\mathbf{E}_{MQW}^{2\omega_1}$ in the MQW structure, and the polarization current $\mathbf{J}_{MQW}^{2\omega_1}$ in the MQW producing an electric field $\mathbf{E}_{FS}^{2\omega_1}$ in the far-field region. In order to determine the electric field radiated by the MQW, we relate the surface current $\mathbf{J}_{(2D)}^{2\omega_1}$ to the electric field of a normally incident plane wave by

$$\mathbf{J}_{(2D)}^{\omega_1} = \frac{2\mathbf{E}_{inc}^{2\omega_1}}{\eta}, \tag{8}$$

where $\eta$ is the free-space impedance. Combining Eqs. (6), (7), and (8) provides the electric field generated by the nonlinear unit-cell as

$$E_{[mab]}^{2\omega_1} = j2k_1\frac{1}{S}\int_{V_{UC}}\chi_{zzz}^{(2)}(\mathbf{r})E_{z[a]}^{\omega_1}(\mathbf{r})E_{z[b]}^{\omega_1}(\mathbf{r})\left(\frac{E_{z[m]}^{2\omega_1}(\mathbf{r})}{E_{inc[m]}^{2\omega_1}}\right)dV, \tag{9}$$

where $k_1$ is the free-space wavenumbers at $\omega_1$, and $E_{[mab]}^{2\omega_1}$ is the $m$-polarized electric field (with $m=\{x,y\}$) generated at $2\omega_1$ by a unit-cell excited by the normally incident fields $E_{inc[a]}^{\omega_1}$ and $E_{inc[b]}^{\omega_1}$. It is important to remark that the proposed approach allows modelling a nonlinear process by solving the three linear electromagnetic problems that provide the fields $E^{\omega_1}_{z[a]}$, $E^{\omega_1}_{z[b]}$, $E^{2\omega_1}_{inc[m]}$, and $E^{2\omega_1}_{z[m]}$ involved in Eq. (9), and that this procedure is accurate for electrically thin nonlinear periodic metasurfaces illuminated at normal incidence with low power conversion efficiencies.

Given that we assume excitation at normal incidence, all elements of the periodic metasurface are excited with the same amplitude and phase, and therefore their radiation will be coherently in phase. Assuming a transverse periodicity smaller than the wavelength, the metasurface will therefore radiate the nonlinear wave towards the normal, and it can be homogenized defining an effective transverse polarizability density

$$\mathbf{P}^{2\omega_1}_{eff} = \frac{\mathbf{J}^{2\omega_1}_{eff}}{j2\omega_1} = \varepsilon_0 \bar{\bar{\chi}}^{(2)}_{eff}\left(I^{\omega_1}\right)\mathbf{E}^{\omega_1}_{inc}\mathbf{E}^{\omega_1}_{inc}, \tag{10}$$

where $\bar{\bar{\chi}}^{(2)}_{eff}\left(I^{\omega_1}\right)$ is the intensity-dependent *effective* nonlinear susceptibility tensor and $\mathbf{J}^{2\omega_1}_{eff}$ is the effective transverse polarization current induced in the material. This current is related to the radiated electric field by

$$\mathbf{J}^{2\omega_1}_{eff} = \frac{2\mathbf{E}^{2\omega_1}}{\eta d}, \tag{11}$$

where *d* takes into account the thickness of the structure. Combining Eqs. (9), (10) and (11) allows to compute each term of the *effective* transverse nonlinear susceptibility tensor of the entire metasurface as

$$\chi^{(2)}_{eff[mab]}\left(I^{\omega_1}\right) = \frac{1}{V}\int_{V_{UC}} \chi^{(2)}_{zzz}(\mathbf{r}) \frac{E^{\omega_1}_{z[a]}(\mathbf{r})}{E^{\omega_1}_{inc[a]}} \frac{E^{\omega_1}_{z[b]}(\mathbf{r})}{E^{\omega_2}_{inc[b]}} \frac{E^{2\omega_1}_{z[m]}(\mathbf{r})}{E^{2\omega_1}_{inc[m]}} dV \ . \tag{12}$$

The importance of Eq. (12) is threefold. First, it demonstrates that each component of the transverse *effective* susceptibility tensor can be engineered to boost the intrinsic nonlinear response of the MQW [$\bar{\bar{\chi}}^{(2)} = \chi_{zzz}^{(2)} \hat{e}_z \hat{e}_z$] by controlling the $E_z$ field enhancement within the cell. Second, it inherently takes phase-matching conditions into account and confirms that a strong overlap of the modal fields is required to enhance the cell nonlinear response. And third, it provides a powerful homogeneous model of complex nonlinear composite structures, including saturation effects arising from the nonuniform intensity distribution. The vertical (*z*-components) of the *effective* susceptibility tensor are strictly zero here, since they do not contribute to the radiation towards the normal, but they may be non-negligible for oblique incidence or for nonperiodic surfaces. Note that this model assumes a spatially local carrier density, neglecting the lateral diffusion that might occur through the MQWs.

The availability of the effective susceptibility tensor $\bar{\bar{\chi}}_{eff}^{(2)}$ allows calculating the metasurface SHG efficiency and generated intensity assuming any arbitrarily polarized pump as

$$SHG_{effic[m]}(I^{\omega_1}) = \frac{1}{2}\eta I^{\omega_1} k_1^2 d^2 \left| \mathbf{e}^{2\omega} \bar{\bar{\chi}}_{eff}^{(2)}(I^{\omega_1}) \mathbf{e}^{\omega} \mathbf{e}^{\omega} \right|^2, \tag{13}$$

$$I^{2\omega_1} = SHG_{effic[m]}(I^{\omega_1}) \cdot I^{\omega_1}, \tag{14}$$

where $\mathbf{e}^{2\omega} = \mathbf{e}_m$ and $\mathbf{e}^{\omega}$ are the polarization unit vectors of the SH and pump beams, respectively. For the sake of compactness, and without loss generality, in what follows we focus on linearly polarized pumps, thus allowing to simplify the conversion efficiency to

$$SHG_{effic[m]}(I^{\omega_1}) = \frac{1}{2}\eta I^{\omega_1} k_1^2 d^2 \left| \chi_{ef[mab]}^{(2)} \right|^2. \tag{15}$$

Importantly, these equations are similar to conventional SHG formulas in nonlinear crystals assuming phase-matched conditions. There is no need to worry about phase matching at the macroscopic level here, since the nonlinearity is concentrated and controlled on a subwavelength scale.

Finally, our technique can be extended to consider nonlinear unit cells simultaneously operating in reflection and transmission. For this purpose, we introduce the generalized susceptibility tensor

$$\ddot{\chi}_{eff}^{(2)} = \begin{pmatrix} \bar{\bar{\chi}}_{eff_{11}}^{(2)} & \bar{\bar{\chi}}_{eff_{12}}^{(2)} \\ \bar{\bar{\chi}}_{eff_{21}}^{(2)} & \bar{\bar{\chi}}_{eff_{22}}^{(2)} \end{pmatrix}, \tag{16}$$

where the subscripts *1* and *2* denote the first and second ports of the structure, i.e., the free-space regions located above and below the unit cells, respectively. In our notation, the first subscript is related to the output port. This port should be excited in a linear electromagnetic simulation to obtain the $E_z$ electric field enhancement within the MQW at the generated frequency, i.e., $\dfrac{E_{z[m]}^{2\omega_1}}{E_{inc[m]}^{2\omega_1}}$ in Eq. (12), required to determine the different susceptibility tensor elements. Similarly, the second subscript is related to the input port, which provides the electric field enhancement within the MQW at the pump frequency, i.e., $\dfrac{E_{z[a]}^{\omega_1}}{E_{inc[a]}^{\omega_1}}\dfrac{E_{z[b]}^{\omega_1}}{E_{inc[b]}^{\omega_1}}$ in Eq. (12). In the special case of cells operating in reflection $\bar{\bar{\chi}}_{eff_{12}}^{(2)} = \bar{\bar{\chi}}_{eff_{21}}^{(2)} = \bar{\bar{\chi}}_{eff_{22}}^{(2)} = 0$.

### *IIb. Nonlinear response for finite-size excitation*

Let us consider that a Gaussian beam with total power $P_0$ is impinging on a SHG plasmonic metasurface. The position-dependent intensity on top of the structure is given by

$$I(\mathbf{r}) = I_0 e^{\frac{-2r^2}{w^2}}, \qquad (17)$$

where $w$ is the beam radius, $\mathbf{r}$ is the position vector within the metasurface and $r = |\mathbf{r}|$. In addition, the intensity peak $I_0$ is related to the input power by

$$I_0 = \frac{2P_0}{\pi w^2}. \qquad (18)$$

Assuming that the unit cells are smaller than the input wavelength, the intensity of the generated beam at each point can be approximated by

$$I^{2\omega}(\mathbf{r}) = I^{\omega}(\mathbf{r}) SHG_{effic[m]}\left(I^{\omega}(\mathbf{r})\right), \qquad (19)$$

thus allowing to compute the total generated power at $2\omega_1$ as

$$P^{SHG} = \iint I^{2\omega_1}(\mathbf{r}) ds. \qquad (20)$$

Then, the overall power conversion efficiency of the nonlinear process yields

$$\eta_{effic}^{SHG} = \frac{P^{SHG}}{P_0^{\omega}}. \qquad (21)$$

This approach assumes that the Gaussian beamwidth is larger than the wavelength, thus allowing to approximate the impinging beam by plane waves of different power normally impinging on the metasurface. In doing so, we take advantage of the weak dependence of the nonlinear metasurface response with the angle of incidence (see [14]).

### III. Saturation effects

Saturation in MQWs occurs when the input intensity is so large that the intersubband absorption empties the quantum-well ground state [2], [36], therefore limiting the overall nonlinear process.

In the case of bulk MQW nonlinear structures, the influence of saturation has usually been modelled [24], [25] by considering i) the variation of the subband population versus the intensity of the incident beam, and ii) the beam propagation in the nonlinear material. The combination of these two factors leads to complex propagation models, and therefore saturation effects have been only evaluated numerically to date. However, this general scenario is simplified in the special case of metasurfaces, whose electrically ultrathin nature prevents any significant propagation of the impinging beams through the MQWs. This fact permits deriving analytical expressions for the saturation factor in nonlinear metasurfaces, while providing physical insight into the nature of the process.

We start our analysis by considering the *zzz*-component of the intersubband nonlinear susceptibility of a SHG nonlinear process at pump frequencies close to the intersubband resonance, which is given by [24]

$$\chi^{(2)}_{zzz}(I_z^{\omega_1}) \approx S^{SHG}(I_z^{\omega_1}) N_e \frac{q_e^3 z_{12} z_{13} z_{23}}{\varepsilon_0 \left(E - E_{12} - i\hbar\gamma_{12}\right)\left(2E - E_{13} - i\hbar\gamma_{13}\right)}, \tag{22}$$

where $-q_e$ is the electron charge, $\hbar$ is the reduced Planck constant, $N_e$ is the averaged bulk density doping, $E = \hbar\omega_1$ is the input energy with $\omega_1$ being the pump frequency, $I_z^{\omega_1}$ is the intensity of z-polarized field at the pump frequency, and $E_{ij}$, $q_e z_{ij}$ and $\hbar\gamma_{ij}$ are the energy, dipole moment and linewidth for transitions between levels *i* and *j*, respectively. The susceptibility depends on the input intensity as a function of the saturation factor $S^{SHG}(I_z^{\omega_1}) \approx (N_1 - 2N_2 + N_3)/N_e$, where $N_i$ is the relative population of the subband *i*. Importantly, this intensity depends solely on the z-component of the electric field within the MQW at the pump frequency, i.e., $I_z^{\omega_1} = 2n_{MQW}^{\omega_1} c \varepsilon_0 \left|E_z^{\omega_1}\right|^2$ [36], where $n_{MQW}^{\omega_1}$ is the MQW refractive index at $\omega_1$.

Saturation effects are negligible $\left[ S^{SHG}(I_z^{\omega_1}) \approx 1 \right]$ when the input intensity is low and therefore the electron population stays predominantly in the ground state, i.e. $N_1 \approx N_e$. However, as the input intensity increases the population of different subbands varies, modifying the features of the nonlinear process. Considering a three-level system, these populations can be approximated from the steady-state solution of the coupled-rate equations [24]

$$\frac{\partial N_1}{\partial t} = -\frac{\alpha_{12}(\omega_1) I_z^{\omega_1}}{\hbar \omega_1} - \frac{\alpha_{13}(2\omega_1) I_z^{2\omega_1}}{\hbar 2\omega_1} + \frac{N_2 - N_{20}}{\tau_{12}} + \frac{N_3 - N_{30}}{\tau_{13}} \to 0, \qquad (23)$$

$$\frac{\partial N_2}{\partial t} = \frac{\alpha_{12}(\omega_1) I_z^{\omega_1}}{\hbar \omega_1} - \frac{\alpha_{23}(\omega_1) I_z^{\omega_1}}{\hbar \omega_1} - \frac{N_2 - N_{20}}{\tau_{12}} + \frac{N_3 - N_{30}}{\tau_{23}} \to 0, \qquad (24)$$

where $N_1 + N_2 + N_3 = N_e$, $\tau_{ij}$ is the relaxation time between subbands $i$ and $j$, $I_z^{2\omega_1}$ is the intensity of the generated second-harmonic, $N_{i0}$ is the carrier concentration in the subband $i$ in thermal equilibrium, the absorption coefficient $\alpha_{ij}(\omega_p)$ between subbands $i$ and $j$ at $\omega_p$ is [24]

$$\alpha_{ij}(\omega_p) = \frac{N_i - N_j}{N_e} \alpha_{ij}^{(0)}(\omega_p) = \frac{(N_i - N_j) \omega_p q_e^2 z_{ij}^2}{\varepsilon_0 n c \left[ (E - E_{ij})^2 + (\hbar \gamma_{ij})^2 \right]}, \qquad (25)$$

and the saturation intensity reads

$$I_{S_{ij}}^{\omega_p} = \frac{N_e \hbar \omega_p}{2 \alpha_{ij}^{(0)}(\omega_p) \tau_{ij}}. \qquad (26)$$

In the following, we assume that the Fermi level is relatively close to the first subband and we neglect the possible influence of optical heating, so $N_{i0} \approx 0$, for $i>1$. Note that our approach is valid for nonlinear processes with low power conversion efficiencies, allowing us to simplify Eq. (23) by neglecting the influence of the generated second-harmonic intensity, i.e., $I_z^{2\omega_1} \approx 0$.

Thanks to the ultrathin nature of metasurfaces, Eqs. (23) and (24) can be solved without considering their coupling to propagation phenomena, yielding to a saturation factor

$$S^{SHG}(I_z^{\omega_1}) \approx \frac{2I_z^{\omega_1}\left[I_{S_{12}}^{\omega_1}(1+\tau_{12}/\tau_{13})-I_{S_{23}}^{\omega_1}(1+\tau_{23}/\tau_{13})\right]+4I_{S_{12}}^{\omega_1}I_{S_{23}}^{\omega_1}(1+\tau_{23}/\tau_{13})}{3\left(I_z^{\omega_1}\right)^2+2I_z^{\omega_1}\left[I_{S_{12}}^{\omega_1}(1+\tau_{12}/\tau_{13})+2I_{S_{23}}^{\omega_1}(1+\tau_{23}/\tau_{13})\right]+4I_{S_{12}}^{\omega_1}I_{S_{23}}^{\omega_1}(1+\tau_{23}/\tau_{13})}. \quad (27)$$

As expected, saturation tends to unity and to zero for very low and high intensities, respectively. It is indeed instructive to study this factor as a function of the features of the third subband. In case that this level barely influences the MQW response, due to a negligible absorption between levels 2 and 3 (i.e., $\alpha_{23} \to 0$ and $N_3 \to 0$), the saturation factor simplifies to the one of a 2 level system

$$S^{SHG}(I_z^{\omega_1}) \approx \frac{N_1-2N_2}{N} = \frac{1}{2}\frac{2I_{S_{12}}^{\omega_1}-I_z^{\omega_1}}{I_{S_{12}}^{\omega_1}+I_z^{\omega_1}}, \quad (28)$$

which is clearly inaccurate for high-input intensities. In the opposite case, considering that the second and third level are strongly coupled (i.e., $\alpha_{23} \to \infty$ and $N_3 \approx N_2$), the saturation factor yields

$$\lim_{\alpha_{23}^{(0)} \to \infty} S^{SHG}(I_z^{\omega_1}) \approx \frac{4I_{S_{12}}^{\omega_1}}{4I_{S_{12}}^{\omega_1}+3I_z^{\omega_1}}. \quad (29)$$

The availability of Eq. (28) and Eq. (29) delimits the overall influence of the third level in the saturation effects, providing a range of well-defined possible solutions uniquely related to the absorption and transitions between first and second levels.

The previous development was focused on saturation in SHG metasurfaces. However, it can easily be extended to consider other nonlinear processes such as DFG, sum frequency generation, etc.

**IV. Limit and potentials of non-linear MQW-based metasurfaces**

In this section, we analytically demonstrate that saturation imposes fundamental limits to the maximum intensity and conversion efficiency that can be attained from nonlinear metasurfaces. These limits depend on the intrinsic MQW properties and the plasmonic nanoresonators design, and our calculations indicate that they can be pushed well above 1% in case of optimized structures. We also determine the intensity of the input beam required to achieve the maximum performance of the metasurface. Let us first consider a nonlinear SHG unit-cell neglecting saturation effects. In this case, the effective nonlinear susceptibility of the structure is given by [14]

$$\chi^{(2)}_{NSeff[mab]} = \frac{\chi^{(2)}_{zzz}}{V} \int_{V_{UC}} \frac{E^{\omega_1}_{z[a]}(\mathbf{r})}{E^{\omega_1}_{inc[a]}} \frac{E^{\omega_1}_{z[b]}(\mathbf{r})}{E^{\omega_1}_{inc[b]}} \frac{E^{2\omega_1}_{z[m]}(\mathbf{r})}{E^{2\omega_1}_{inc[m]}} dV = \chi^{(2)}_{zzz} I^{overlap}_{[mab]}, \quad (30)$$

where the subscript "NS" stands for "neglecting saturation" and $I^{overlap}_{[mab]}$ is the modal overlapping integral for a given polarization. There are many physical configurations that, although possessing different electric field distributions within the unit-cell, provide the same $\bar{\bar{\chi}}^{(2)}_{NSeff}$ susceptibility in the absence of saturation. These unit cells might use different geometries to implement the plasmonic resonator, combine various metals and dielectrics, or employ differently engineered MQWs. Ideally, all these cases would lead to metasurfaces with identical nonlinear response. When saturation effects are taken into account, however, the intrinsic field distribution of each particular unit cell plays a major role. Considering an impinging electromagnetic wave, saturation will mainly attenuate the contributions to the effective nonlinear susceptibility from the MQW regions where the induced intensity is very large, whereas it will barely affect the contributions from the areas with little induced intensity. Consequently, the overall intensity-dependent effective nonlinear response $\bar{\bar{\chi}}_{eff}^{(2)}$ of different metasurfaces with identical $\bar{\bar{\chi}}_{NSeff}^{(2)}$ can significantly vary. This raises an important question: what is the optimal nonlinear performance achievable by a metasurface composed by cells with a given $\bar{\bar{\chi}}_{NSeff}^{(2)}$? As previously pointed out, this performance is closely related to the $E_z$ field distribution within the MQW. Carefully analysing Eq. (12) and considering the $\propto \dfrac{1}{I_z^{\omega_1}}$ dependence of the saturation factor, it is easy to demonstrate that an optimal unit-cell would provide a constant $E_z$ field enhancement over the whole volume at $\omega_1$, i.e. $F_{e[ab]}^{\omega 1} = \dfrac{E_{z[a]}^{\omega_1}}{E_{inc[a]}^{\omega_1}} = \dfrac{E_{z[b]}^{\omega_1}}{E_{inc[b]}^{\omega_1}}$. This allows simplifying the components of $\bar{\bar{\chi}}_{NSeff}^{(2)}$ to

$$\chi_{NSeff[mab]}^{(2)} = \chi_{zzz}^{(2)} F_{e[m]}^{2\omega_1} \left(F_{e[ab]}^{\omega_1}\right)^2, \tag{31}$$

where $F_{e[m]}^{2\omega_1}$ is the $E_z$ field enhancement within the MQW at $2\omega_1$. This "ideal" cell minimizes the influence of the saturation process, and therefore provides a theoretical upper bound for the performance of any nonlinear unit-cell described by the same $\bar{\bar{\chi}}_{NSeff}^{(2)}$. The effective intensity-dependent susceptibility tensor of this "ideal" cell is given by

$$\bar{\bar{\chi}}_{eff_B}^{(2)}\left(I^{\omega_1}\right) = \bar{\bar{\chi}}_{NSeff}^{(2)} S\left(I_z^{\omega_1}\right), \tag{32}$$

where the intensity $I_z^{\omega_1}$ induced in the structure is related to the intensity $I^{\omega_1}$ of the impinging waves by [2]

$$I_z^{\omega_1} = I^{\omega_1} n_{MQW}^{\omega_1} \left(F_{e[ab]}^{\omega_1}\right)^2. \tag{33}$$

Considering an incident beam with high intensity $I^{\omega_1} \gg 2/3(I_{S1} + 2I_{S2})$, the effective susceptibility can be simplified to

$$\chi^{(2)}_{eff[mab]}\left(I^{\omega_1}\right) \approx \frac{2}{3}\chi^{(2)}_{zzz}F^{2\omega_1}_{e[m]}\frac{I^{\omega_1}_{S_{12}}\left(1+\tau_{12}/\tau_{13}\right)-I^{\omega_1}_{S_{23}}\left(1+\tau_{23}/\tau_{13}\right)}{n^{\omega_1}_{MQW}I^{\omega_1}}, \tag{34}$$

which solely depends on the field enhancement at the generated frequency $F^{2\omega_1}_{e[m]}$ and on the characteristics of the MQW, including saturation currents and permittivity. As expected, the SHG efficiency of the unit cell tends to zero for input beams with extremely high intensity, i.e.,

$$\lim_{I^{\omega_1}\to\infty} SHG_{effic}(I^{\omega_1}) \approx 0, \tag{35}$$

whereas the maximum value of the generated intensity is upper-bounded to

$$\lim_{I^{\omega_1}\to\infty} I^{2\omega_1}\left(I^{\omega_1}\right) = \frac{2}{9}\eta k_1^2 \frac{d^2}{n^2_{MQW}}\left|\chi^{(2)}_{zzz}F^{2\omega_1}_{e[m]}\right|^2\left[I^{\omega_1}_{S_{12}}\left(1+\tau_{12}/\tau_{13}\right)-I^{\omega_1}_{S_{23}}\left(1+\tau_{23}/\tau_{13}\right)\right]^2. \tag{36}$$

Remarkably, the maximum generated intensity does not depend on the field enhancement at the pump frequency. In addition, the intensity generated at $2\omega_1$ may present a local maximum for input beams of intensity

$$I^{\omega_1}_M = \frac{2I^{\omega_1}_{S_{12}}I^{\omega_1}_{S_{23}}\left(1+\tau_{23}/\tau_{13}\right)}{n^{\omega_1}_{MQW}\left(F^{\omega_1}_{e[ab]}\right)^2\left[I^{\omega_1}_{S_{23}}\left(1+\tau_{23}/\tau_{13}\right)-I^{\omega_1}_{S_{12}}\left(1+\tau_{12}/\tau_{13}\right)+\sqrt{3I^{\omega_1}_{S_{23}}\left(1+\tau_{23}/\tau_{13}\right)\left(I^{\omega_1}_{S_{23}}\left(1+\tau_{23}/\tau_{13}\right)-I^{\omega_1}_{S_{12}}\tau_{12}/\tau_{13}\right)}\right]},$$

(37)

which exists only if a positive real solution of the above equation exists. Furthermore, it is indeed useful to determine the intensity of input beams that provides the maximum conversion efficiency, $I^{\omega_1}_C$. This can be done by solving the equation

$$\frac{\partial}{\partial I^{\omega_1}}\left[S\left(I^{\omega_1}_z\right)I^{\omega_1}\right] = 0, \tag{38}$$

which has a rather complicated solution, not shown here for the sake of brevity. Its expression can be greatly simplified considering either a 2-level system or a strong coupling between the second and third subbands. Solving these cases analytically, the required input intensity is found to be in the range of

$$\frac{(\sqrt{33}-5)}{2} \frac{I_{S_{12}}^{\omega_1}}{n_{MQW}^{\omega_1} \left(F_{e[ab]}^{\omega_1}\right)^2} \leq I_C^{\omega_1} \leq \frac{4}{3} \frac{I_{S_{12}}^{\omega_1}}{n_{MQW}^{\omega_1} \left(F_{e[ab]}^{\omega_1}\right)^2}, \tag{39}$$

where the lower bound of $I_C^{\omega_1}$ is being determined by considering a 2-level system and the upper bound is being determined by considering strong coupling between the second and third subbands. The range of maximum SHG conversion efficiencies then becomes:

$$\left(\frac{\sqrt{33}-5}{4}\right)\left(\frac{9-\sqrt{33}}{\sqrt{33}-3}\right)^2 C_{eff} \leq SHG_{eff}^{MAX} \leq \frac{2}{3} C_{eff}, \tag{40}$$

where

$$C_{eff} = \frac{\eta k_1^2 d^2}{4} \frac{I_{S_{12}}^{\omega_1}}{n_{MQW}^{\omega_1}} \left|F_{e[m]}^{2\omega_1}\right|^2 \left|F_{e[ab]}^{\omega_1}\right|^2 \left|\chi_{zzz}^{(2)}\right|^2. \tag{41}$$

Contrary to the case of the generated intensity, the maximum conversion efficiency indeed depends on the $E_z$ field enhancement of the structure at the pump frequency. In summary, the availability of Eqs. (34)-(41) permits determining the fundamental limits imposed by saturation over any nonlinear unit-cell with a given $\bar{\bar{\chi}}_{NSeff}^{(2)}$, including i) maximum generated intensity and conversion efficiency, and ii) intensity of the input beam required to achieve these values. More importantly, it also provides the required guidelines to design structures with desired nonlinear behaviour.

The analysis described above has profound implications for the design of practical nonlinear metasurfaces, even in case that they do not present a uniform $E_z$ field enhancement within the unit-cells. In fact, the linear nature of the effective intensity-dependent susceptibility [see Eq. (12)

] of any unit-cell permits its computation by simply adding the susceptibilities of its composing regions. Then, considering regions small enough to provide a constant $E_z$ field enhancement at the pump and generation frequencies, one can determine the fundamental limits of each region. The analytical nature of these limits provides the required guidelines to design non-linear metasurfaces with optimized performance. Finally, note that this approach can easily be generalized to derive the fundamental limits of any other nonlinear process.

## V. Experimental validation

In this section, we validate the numerical approach derived above using experimental data. For this purpose, we analyse the plasmonic metasurface with giant nonlinearities recently reported in [14]. This reference also describes in detail the experiment, including sources, detectors, calibration procedures, etc. We have confirmed that our model works well for other fabricated MQW-based nonlinear metasurfaces, not included here for the sake of compactness. The unit cell that composes the device under analysis is illustrated in the inset of Fig. 2a. It is essentially composed of an InGaAs/AlIn MQW structure with theoretical $\chi_{zzz}^{(2)} \approx 50 nmV^{-1}$ and saturation intensity $I_{S12}^{\omega} \approx 0.47 MWcm^{-2}$ sandwiched between a metallic ground plane and a plasmonic resonator, providing SHG in reflection. We assume a similar saturation intensity for transitions between levels 2 and 3, which is justified by the symmetry of the MQW structure.

Fig. 2a shows the measured *x*-polarized SHG beam power generated by the metasurface versus the square of the power of an impinging *x*-polarized Gaussian beam with radius of $17 \mu m$. The pump and SHG frequencies are 37 and 74 THz, respectively. Fig. 2b depicts similar measurements, but for *y*-polarized incident and reflected beams. Our numerical analysis is able to fit measured data using effective nonlinear susceptibilities of $\chi^{(2)}_{NSeff[xxx]} = 36.5 nm/V$ and $\chi^{(2)}_{NSeff[yyy]} = 51.8 nm/V$ for *x* and *y*-polarized, respectively, incident and reflected fields. This allows to indirectly measure an intrinsic MQWs susceptibility of $\chi^{(2)}_{zzz} \approx 95\, nmV^{-1}$, which is in relatively good agreement with its theoretical value given the uncertainties of the experimental setup [14]. Simulation results, computed neglecting the influence of saturation (blue dashed-line), shows the expected linear dependence of the SH power versus the square the input power. However, this dependence is only accurate for input beams with very low power. As this power increases, the saturation of the intersubband transitions in the MQW structure significantly affects the SHG power, changing the slope of the curve and limiting the metasurface performance. This phenomenon is rigorously taken into account by the theory developed in the present work. The conversion efficiency of the metasurface for these two cases is shown in Fig. 2c. Our simulation results (red dashed-lines in Fig. 2) show an excellent agreement with measurements, fully confirming the accuracy of the proposed technique to characterize the nonlinear response of plasmonic metasurfaces. This comparison further confirms the accuracy of our spatially local carrier density model, and suggests that diffusion phenomena does not really play a dominant role in nonlinear metasurfaces, but rather seems to be a second-order effect that can be neglected in a first approximation. We stress that the fit with measurements is obtained without fitting parameters, but directly based on our analytical saturation model.

In order to fully understand the behaviour of this device, we apply the theory developed above to numerically investigate its nonlinear response considering input beams of high power. Fig. 3a shows the *y*-polarized SHG intensity and conversion efficiency of the unit cell versus the intensity of incident *y*-polarized light. In case of low intensity, i.e., when the influence of saturation is limited, SHG intensity and conversion efficiency increase with the intensity of the input beams. The unit-cell provides a maximum conversion efficiency of $2.7 \cdot 10^{-4}$% for incident fields with intensity of 28.2kW/cm$^{-2}$, and a maximum output intensity of 0.181W/cm$^{-2}$, achieved for input intensities around 143kW/cm$^{-2}$. Then, as the impinging light intensity is further increased, saturation becomes the dominant phenomenon that determines the response of the metasurface, strongly attenuating its nonlinear response. As expected, conversion efficiency tends to zero for input beams with very high power. Interestingly, as predicted by Eq. (37) for cells with uniform field distribution, the SHG intensity seems to follow this trend. This behaviour appears because the input intensities considered in the analysis saturate the nonlinearities from the regions of the unit-cell that present large mode overlapping, but they are not high enough to induce a strong nonlinear response from the areas with low $E_z$ field enhancement. As the input intensity further increases, the SHG intensity grows again, saturating at 1.97W/cm$^{-2}$ for extremely intensive incident light. Fig. 3b shows the *y*-polarized SHG power and conversion efficiency of the entire metasurface considering a *y*-polarized impinging Gaussian beam. Results show a similar overall response and trends as in the case of a single unit-cell.

Finally, we compare the conversion efficiency of the fabricated device with the one of an ideal unit-cell that provides uniform $E_z$ field enhancement across the entire structure, i.e. same enhancement is assumed for fundamental and SHG frequencies, and present a similar effective susceptibility ( $\chi^{(2)}_{NSeff[yyy]} = 51.8\,nm/V$ ), in order to clearly determine the influence of the unit-cell design in the saturation of the generated fields. This ideal cell would provide a maximum conversion efficiency of around $1.3 \cdot 10^{-3}$%, clearly outperforming the initial design. This demonstrates the importance of saturation in the proposed nonlinear plasmonic metasurfaces, and it emphasizes the importance of having uniform field enhancement in the entire MQW structure.

## VI. Improved designs

In this section, we numerically investigate the behaviour of MQW-based nonlinear metasurfaces. First, we provide general guidelines to design these structures, and then we propose a novel unit-cell design to achieve efficient SHG. We also study the performance of the proposed device versus the input intensity and compare them with the response obtained using ideal unit-cells. Finally, we investigate the influence of the MQW design on the nonlinear performance of the device and we demonstrate that an adequate engineering of quantum wells can significantly boost the response of nonlinear metasurfaces. Our study neglects the possible influence of thermal effects, which strongly depends on light pulse duration and well as thermal management of the sample, since the maximum intensity induced in the MQWs is well below the material damage threshold [2].

The proper design of MQW-based plasmonic metasurfaces faces important challenges. First, plasmonic inclusions must be appropriately designed to excite electric fields *perpendicular* to the structure, thus boosting optical transitions within the MQWs. In this sense, the ultimate goal is to maximize the $E_z$ field enhancement in the entire structure, while simultaneously boosting the overlapping modal integral between the difference resonances. In addition, saturation effects will limit the contribution to the nonlinear susceptibility from these nonlinear regions where the induced intensity - closely related to the field enhancement – is larger. This imposes an important trade-off in the design: on one hand, structures with large modal overlap integral [see Eq. (12)] require asymmetric field distribution within the MQW, on the other hand, geometries with uniform $E_z$ field enhancement are less affected by saturation. Furthermore, in case of metasurfaces aiming at maximizing the generated intensity, the plasmonic resonators should boost the $E_z$ field enhancement mostly at the nonlinear frequency. However, if the target is to enhance the maximum conversion efficiency of the nonlinear process, then the field enhancement at the pump and generation frequencies should be equally optimized as follows from Eq. (39). Regarding the unavoidable dissipation losses of the metallic nanoresonators, they slightly diminishes the effective nonlinear susceptibility of the metasurface. The second main challenge is to engineer MQWs able to boost the nonlinear susceptibility, while maximizing the associated saturation currents. In this respect, it is important to note large nonlinear susceptibilities are usually achieved by strongly doping the semiconductor, which in turn increases the dissipation losses of the structure. Therefore, it is required to find a trade-off between the nonlinear response of MQWs and their associated losses.

The nonlinear plasmonic metasurface proposed to achieve efficient SHG is illustrated in Fig. 1. The InGaAs/AlIn MQWs employed in our simulations possess a computed longitudinal optical (LO) phonon scattering decay rate between the different subbands of $\tau_{12} = 1.5\,ps$, $\tau_{13} = 2.45\,ps$ and $\tau_{23} = 1.5\,ps$, and transition dipole moments of $z_{12} = -1.59\,nm$, $z_{13} = -0.88\,nm$, and $z_{23} = -2.76\,nm$, in good agreement with recent measurements [15]. In addition, we consider high-quality semiconductor heterostructures with transition linewidths of $\hbar\gamma_{12} = \hbar\gamma_{13} = \hbar\gamma_{23} \approx 5\,meV$, as in [24],[37], [38]. Fig. 4a shows the intrinsic nonlinear susceptibility $\chi^{(2)}_{zzz}$ of the MQW versus the average doping level $N_e$, confirming that an increase in the doping enhances the overall nonlinear response. The saturation intensity for transitions between the first and second energy levels, illustrated in the inset of Fig.4a, is $I^{\omega_1}_{S_{12}} \approx 24\,MWcm^{-2}$ for the fundamental frequency of 27

THz at which $\chi_{zzz}^{(2)}$ is maximized. We note that this fundamental frequency corresponds to approximately ½ of the transition frequency between states 3 and 1 in the MQW structure, as illustrated in the bandstructure diagram shown in the inset of Fig. 4b. In addition, Fig. 4b also depicts the effective uniaxial permittivity tensor that describes the optical response of the MQWs [2], [14], [24]. The parallel components of the effective permittivity ($\varepsilon_\parallel$) are independent of $N_e$, whereas the perpendicular components ($\varepsilon_\perp$) show a resonant behaviour at the design frequencies. As expected, these resonances are more pronounced as the doping increases, enhancing losses and absorption at the pump and generation frequencies. The unit-cell that compose the proposed metasurface is shown in Fig. 1b, and it comprises a gold T-shaped resonator deposited over the (grounded) MQW. Fig. 4c shows the absorption spectrum of the cell versus the MQW average doping, exhibiting very high absorption for *y* and *x*- polarized waves at 54 THz and 27 THz, respectively. The optimization of the resonator geometry has rigorously considered the $E_z$ mode overlapping among the different resonances as implied by Eq. (12), thus boosting the effective nonlinear susceptibility of the structures. This is clearly illustrated in Fig. 4d, where the real part of the mode overlapping within the MQWs volume is shown. The overlapping is greatly enhanced in the area below the edges of the T-shaped resonator, while small out-of-phase contributions are generated from the central region of the cell. The figure inset depicts the evolution of the overlapping integral [neglecting the influence of saturation, see Eq. (30)], versus the MQW doping. It can be observed that increasing the average doping reduces the modal integral, which is attributed to the increase of the dissipation losses that in turn limit the maximum $E_z$ field enhancement in the structure.

The nonlinear optical response of the proposed metasurface is shown in Fig. 5. The largest element in the effective nonlinear susceptibility tensor of the unit-cell, neglecting saturation effects, is $\chi^{(2)}_{NSeff[yxx]} = 396 nmV^{-1}$. The plasmonic resonance play the fundamental role of coupling the impinging light to the direction perpendicular to the structure, thus fully exploiting and boosting the nonlinear susceptibility $\chi^{(2)}_{zzz}$ of the MQWs. This figure also confirms that the effective nonlinear susceptibility $\chi^{(2)}_{eff}$ (solid red line) saturates fast with the input intensity, i.e., it presents a strong dependence to the induced intensity that yields to weak nonlinear responses even for impinging light of moderate intensity. In addition, Fig. 5a depicts the nonlinear susceptibility of a unit-cell (dashed blue line) with the same $\chi^{(2)}_{NSeff}$ as the original geometry, and an identical uniform $E_z$ field enhancement within the MQWs for fundamental and SHG frequencies (labelled as "ideal unit cell"). As expected, the influence of saturation effects is significantly reduced here. Once the effective susceptibility of the plasmonic unit-cell is known, the performance of the nonlinear metasurface can be easily computed using the theory developed in Section IIb. Specifically, Figs. 5b-c show the generated intensity (b) and output power (c) of the metasurface. The power of the generated light monotonically increases with the power of the impinging beams, allowing the generation of beams with 30 mW of power when the structure is illuminated by beams of 20 W. The excellent response of the ideal structure, also included in the figures, implies that the metasurface performance may be improved around two orders of magnitude further by employing even better optimized resonator geometries able to keep the field uniformity within the MQW. Fig. 5d depicts the conversion efficiency of the structure, achieving 0.2 % for input beams of 6 W of power, all over a deeply subwavelength thickness. This result can be potentially increased up to 1.25% with an "ideal unit cell" with uniform $E_z$ field distribution. It is important to point out that the proposed SHG metasurface outperforms by three orders of magnitude, in both conversion efficiency and maximum generated power, the design presented in [14] and analysed in Section V.

The second main strategy to further enhance the response of nonlinear metasurfaces is to employ MQWs with optimized features. To this purpose, Fig. 6 explores the power conversion efficiency of the proposed device for different parameters of the semiconductor heterostructures. Specifically, Fig. 6a studies the evolution of the conversion efficiency versus the average doping of the MQWs. Simulation results demonstrate that increasing the doping enhances the maximum conversion efficiency and also up-shifts the power of the impinging beams required to achieve it, thus boosting the overall performance of the metasurface. This implies that an increase in doping levels appears overall beneficial to maximize the efficiency, despite its negative effect on absorption and on the modal integrals highlighted above. However, the maximum doping level that can be employed in the design is in practice limited by the Fermi level of the semiconductor. Fig. 6b investigates the influence of the second band energy level in the conversion efficiency of the metasurface. As inferred from Section III, increasing the energy of this band leads to MQWs with lower nonlinear susceptibilities and higher saturation currents. The combination of these two factors permits increasing the input power required to attain the maximum conversion efficiency of the device, which remains relatively constant in all cases. Fig. 6c depicts the dependence of the conversion efficiency versus the linewidth of the quantum system, i.e. $\gamma = \gamma_{12} \approx \gamma_{23} \approx \gamma_{13}$. The minimum attainable linewidth is limited at room temperature by dephasing induced mostly by phonon scattering and at low temperatures by the non-parabolicity of the semiconductor heterostructures [24]. The value of $\gamma$ has a paramount effect on the conversion efficiency of the metasurface, leading to conversion efficiencies that may range from 0.02% to 1.2% for the resonator designs shown in Fig. 1. Finally, Fig. 6d shows the efficiency of the structure versus the relaxation time between the first and second energy levels. As expected, a shorter relaxation time entails higher saturation currents, which in turns enhances the performance of the device.

**VII. Conclusions**

In summary, we have presented a comprehensive theory to analyse the electromagnetic response of MQW-based nonlinear metasurfaces, and validated it with numerical simulations and experimental results. The approach provides a homogeneous effective susceptibility tensor for various nonlinear processes occurring in ultrathin metasurfaces operating either in reflection or in transmission, rigorously taking into account the influence of saturation and losses. Our study has analytically revealed that saturation imposes fundamental limits to the nonlinear response of metasurfaces, and general design rules have been provided to minimize its influence.

This work has established the theoretical foundations for the analysis and design of highly efficient nonlinear periodic metasurfaces, and can be further extended to consider additional scenarios, non-linear processes, and phenomena. For instance, in practical situations non-linear metasurfaces might be excited by pump beam(s) at oblique incidence angles. Our proposed approach can easily analyse this experimental configuration by taking into account the perpendicular ($z$) components of the effective non-linear susceptibility tensor, which are not relevant in case of normal excitation. In addition, future developments of this theory may also consider metasurfaces composed of nonuniform unit-cells with different phase-response for SHG generation. These structures would allow manipulating the generated beams, thus providing exciting functionalities such as focusing or beam steering. Future advances will also allow to treat non-resonant MQWs, which might lead to even larger nonlinearities. Another challenge is to rigorously consider the dependence of the MQW permittivity with respect to the induced intensity due to saturation, which might affect the total $E_z$ field enhancement provided by the plasmonic resonances. Finally, it would also be desirable to combine our electromagnetic macroscopic approach with quantum transport theories able to deal with the diffusion of the carriers in the MQW layers and with thermodynamic techniques able to take thermal effects into account, thus leading to fully multi-physics models.

We need to emphasize that the examples of the unit cell designs shown in this paper are by no means optimal, and we expect to achieve even larger nonlinear responses using refined structures that may provide higher and more uniform $E_z$ field enhancements at fundamental and SHG frequencies, compared to the structures discussed here. To this purpose, novel unit-cells with advanced plasmonic resonator geometries combined with adequately tailored MQWs need to be designed. The optimization of the plasmonic resonance would allow to virtually engineering any element of the *effective* nonlinear susceptibility tensor to achieve a desired operation. In addition, in order to enhance the efficiency of the nonlinear process, this optimization must maximize the mode overlapping among the different resonances of the structure, keeping a uniform $E_z$ field distribution at the pump frequency - in order to minimize the influence of saturation - and boosting the $E_z$ field enhancement at the generation frequency. On the other hand, MQW structures should at the same time provide high nonlinear susceptibility and large saturation intensities, keeping in mind that high nonlinearities are associated with large dissipation losses. In this regard, MQW structures based on other material systems can be explored, such as InAs/AlSb or GaAs/AlGaAs. We envision that the combination of these design approaches will lead to highly efficient nonlinear devices, with power conversion efficiencies exceeding 10%, for low (10-100 kW/cm$^2$) pump intensities and over subwavelength thicknesses, fully confirming the suitability of MQW-based plasmonic metasurfaces to create a novel flatland nonlinear optics paradigm able to generate and manipulate any nonlinear process at a mesoscopic level, with significant impact on a variety of applications such as THz generation/detection, frequency conversion, and phase conjugation.

**Acknowledgments**

This works has been supported by the NSF grant ECCS-1348049, and by the AFOSR grant No. FA9550-14-1-0105.

**Figures**

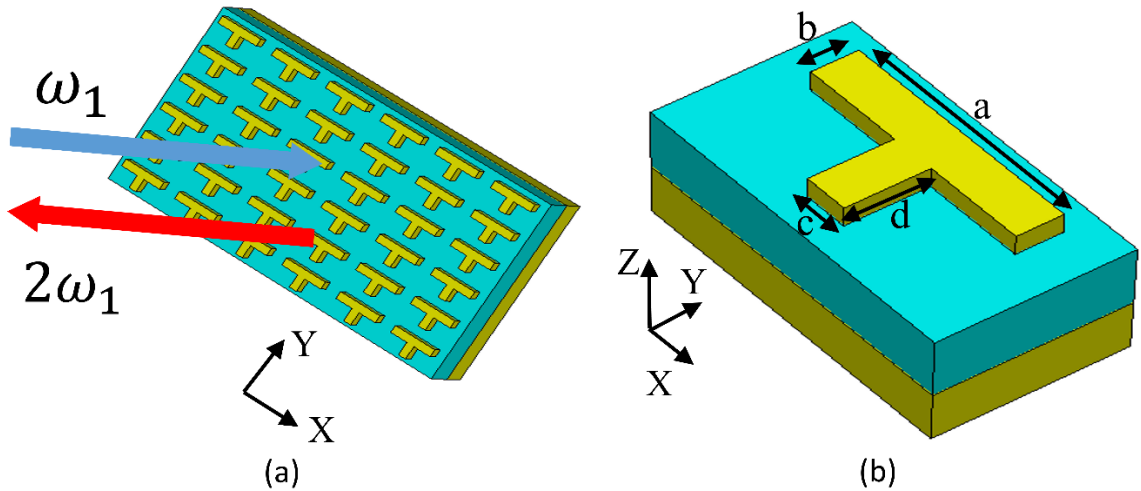

(a)  (b)

Figure 1 – Proposed MQW-based plasmonic metasurfaces for SHG. (a) Schematic overview of the entire metasurface and nonlinear process. (b) Detail of the subwavelength unit-cell composing the metasurface. Geometrical parameters are: a=1.4 um, b=0.25 um, c=0.25 um, and d=0.45 um. Thickness of gold (yellow) and MQW (cyan) are 50 nm and 400 nm, respectively. Our analysis assumes that both impinging and generated beams are in the direction perpendicular to the metasurface.

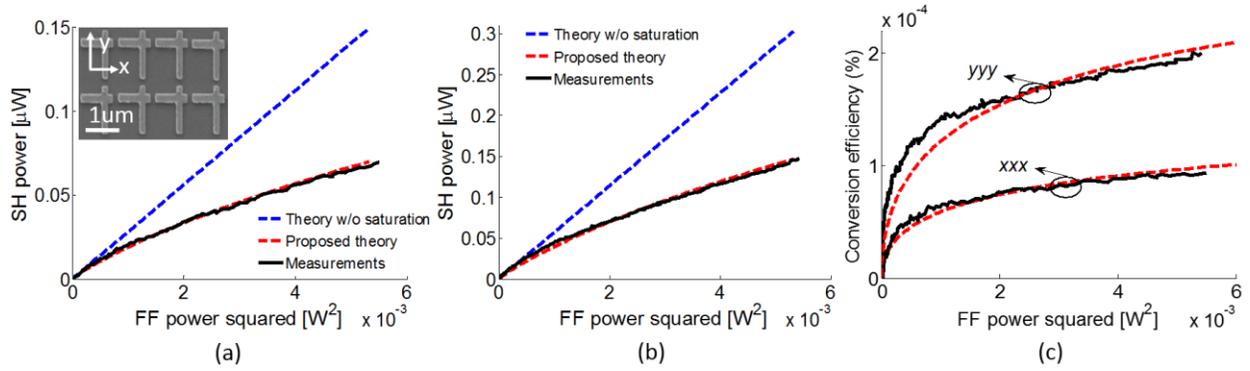

Figure 2 – Generation of second harmonic beams using the nonlinear plasmonic metasurface reported in [14]. The inset in panel (a) shows a picture of the fabricated device. Results are computed with (red dashed-line) and without (blue dashed-line) considering saturation effects. Measurements (black solid-line) are included for comparison purposes. (a) *xxx*-polarization. (b) *yyy*- polarization. (c) Conversion efficiency of panels (a) and (b). The impinging Gaussian beam has a radius of 17 um.

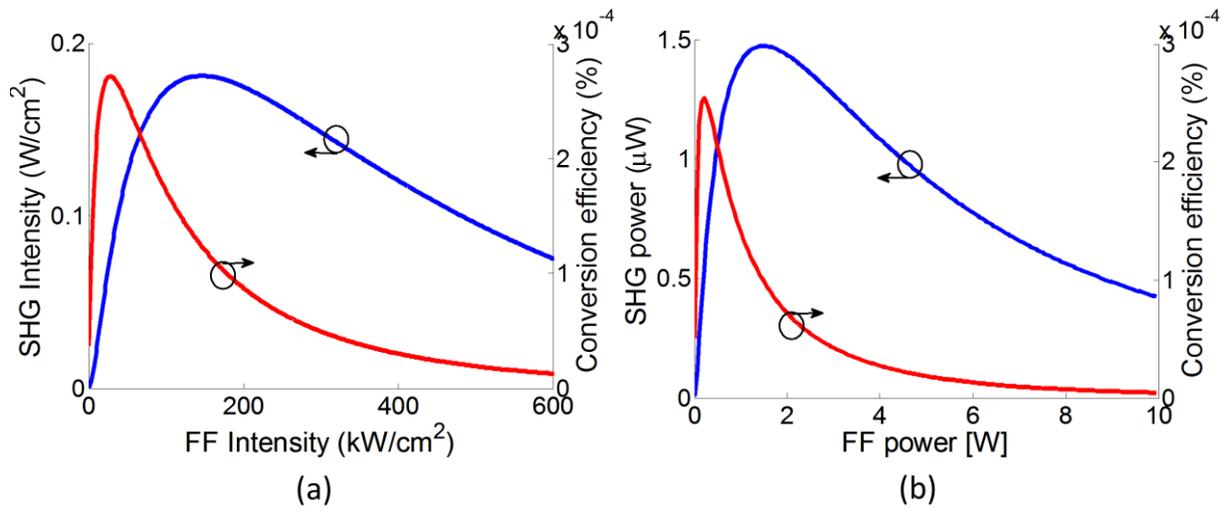

Figure 3 – Simulated *y*-polarized response of the nonlinear plasmonic metasurface reported in [14] excited by *y*-polarized light. Results are computed including saturation effects. (a) SHG intensity and conversion efficiency versus the intensity of the incident fields. (b) SHG power and conversion efficiency versus the power of an incident Gaussian beam (radius of 17 um).

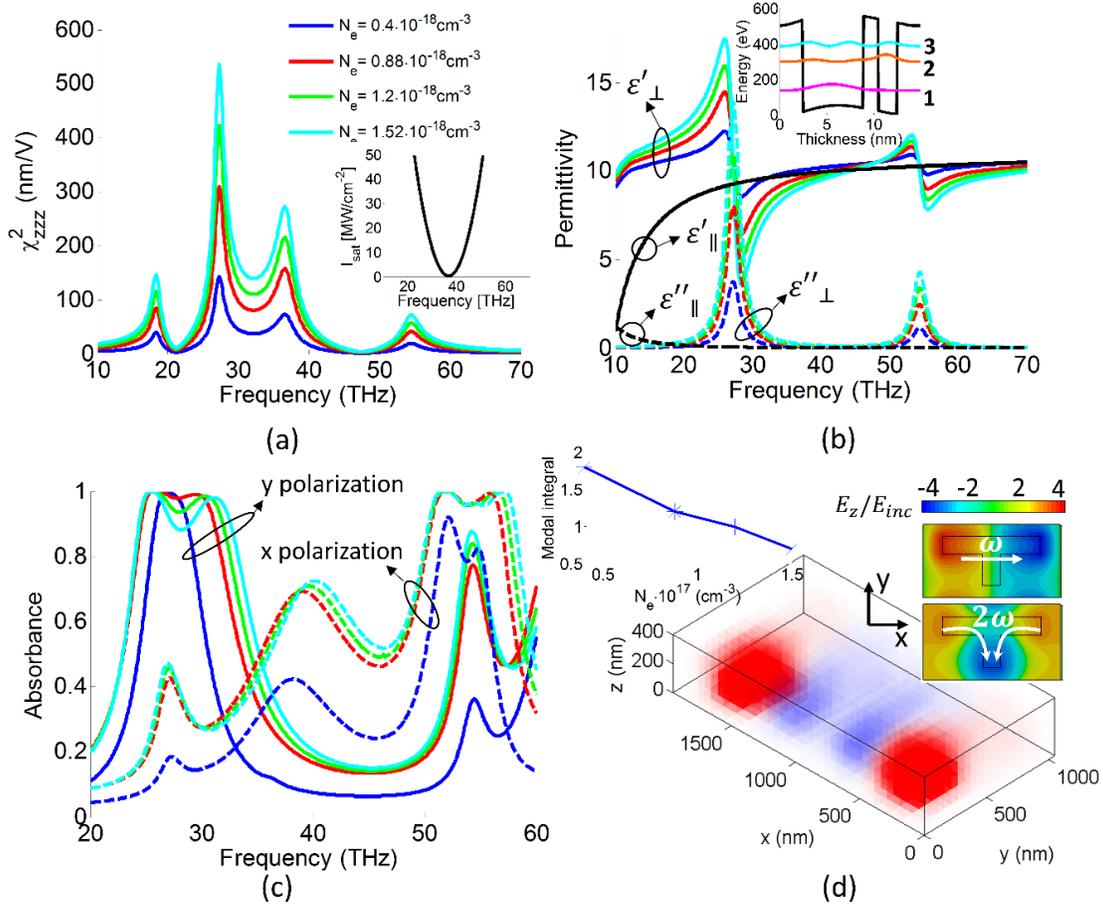

Figure 4 – Influence of MQW doping ($N_e$) on the electromagnetic properties of the nonlinear metasurface shown in Fig. 1. (a) MQW intrinsic nonlinear susceptibility $\chi^{(2)}_{zzz}$. (b) Effective permittivity tensor of the MQWs. The inset shows the conduction band diagram of one period of the designed quantum well structure. The moduli squared of the electron wavefunctions for subbands 1,2 and 3 is also shown. (c) Simulated absorption spectrum of the entire metasurface for different polarizations of the incident light. (d) Real part of the volumetric mode overlapping within the MQWs [see integrant of Eq. (12)]. Saturation effects are not taken into account, i.e., nonlinear susceptibility is assumed to be constant within the whole volume of the unit cell. The insets show (left) overlap integral versus the MQW doping, taking losses into account, and (right) the top view of the normalized $E_z$ field distribution on the cells at the pump and generation frequencies.

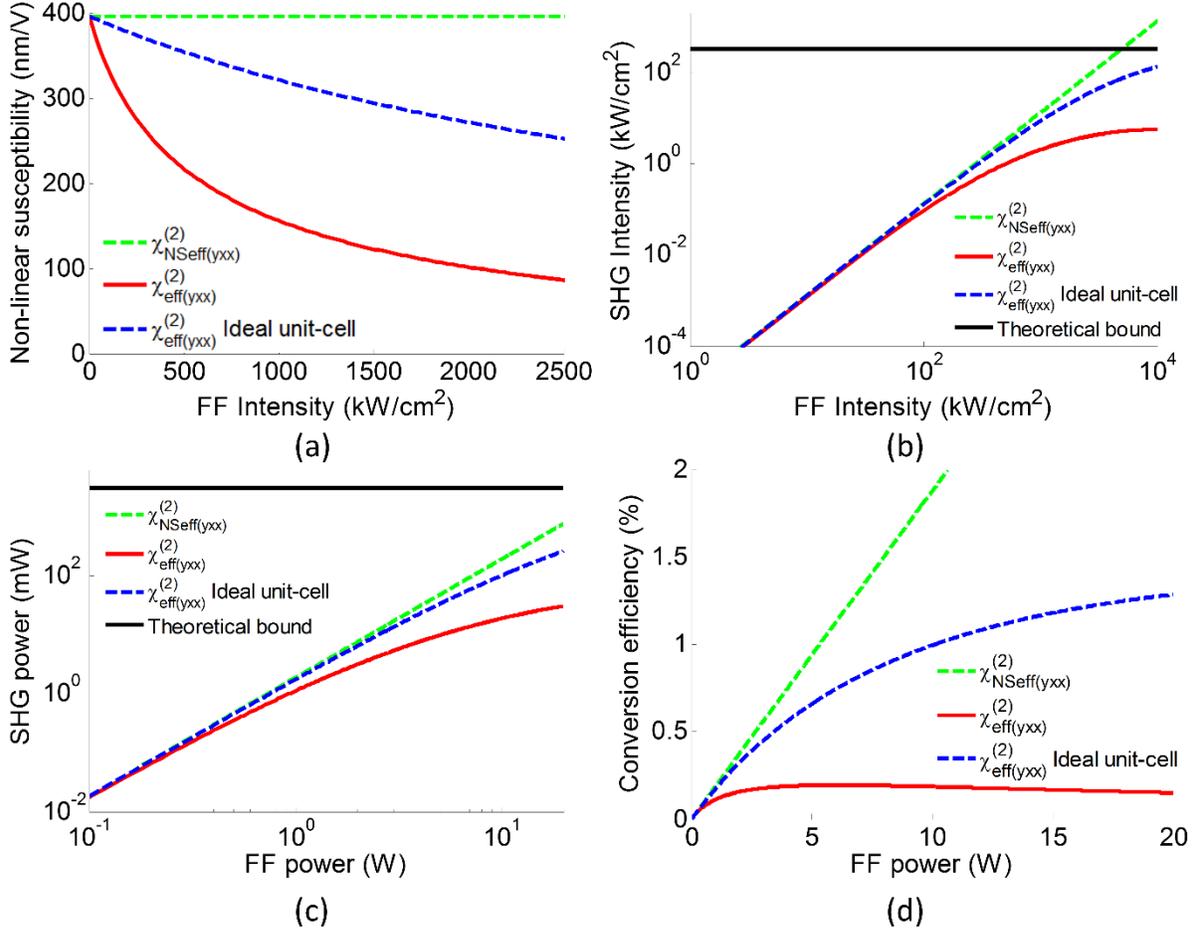

Figure 5 – Influence of saturation on the nonlinear properties of the SHG metasurface shown in Fig. 1. (a) Evolution of the unit-cell effective susceptibility with and without considering saturation, i.e. $\chi^{(2)}_{eff(yxx)}$ (solid line-red) and $\chi^{(2)}_{NSeff(yxx)}$ (dashed line-green), respectively, versus the intensity of the incident light. The dashed blue line represents the intensity-dependent effective nonlinear susceptibility of an ideal unit-cell design that possesses the same nonlinear reponse - $\chi^{(2)}_{NSeff(yxx)}$ - as the original structure. (b) SHG intensity versus the intensity of the incident light. Generated power (c) and conversion efficiency (d) of the proposed metasurface when is excited by an incident *x*-polarized Gaussian beam (17um of radius). MQW doping is set to $N_e = 8.8 \cdot 10^{-17}$ cm$^{-3}$. The inset in panel b shows the energy

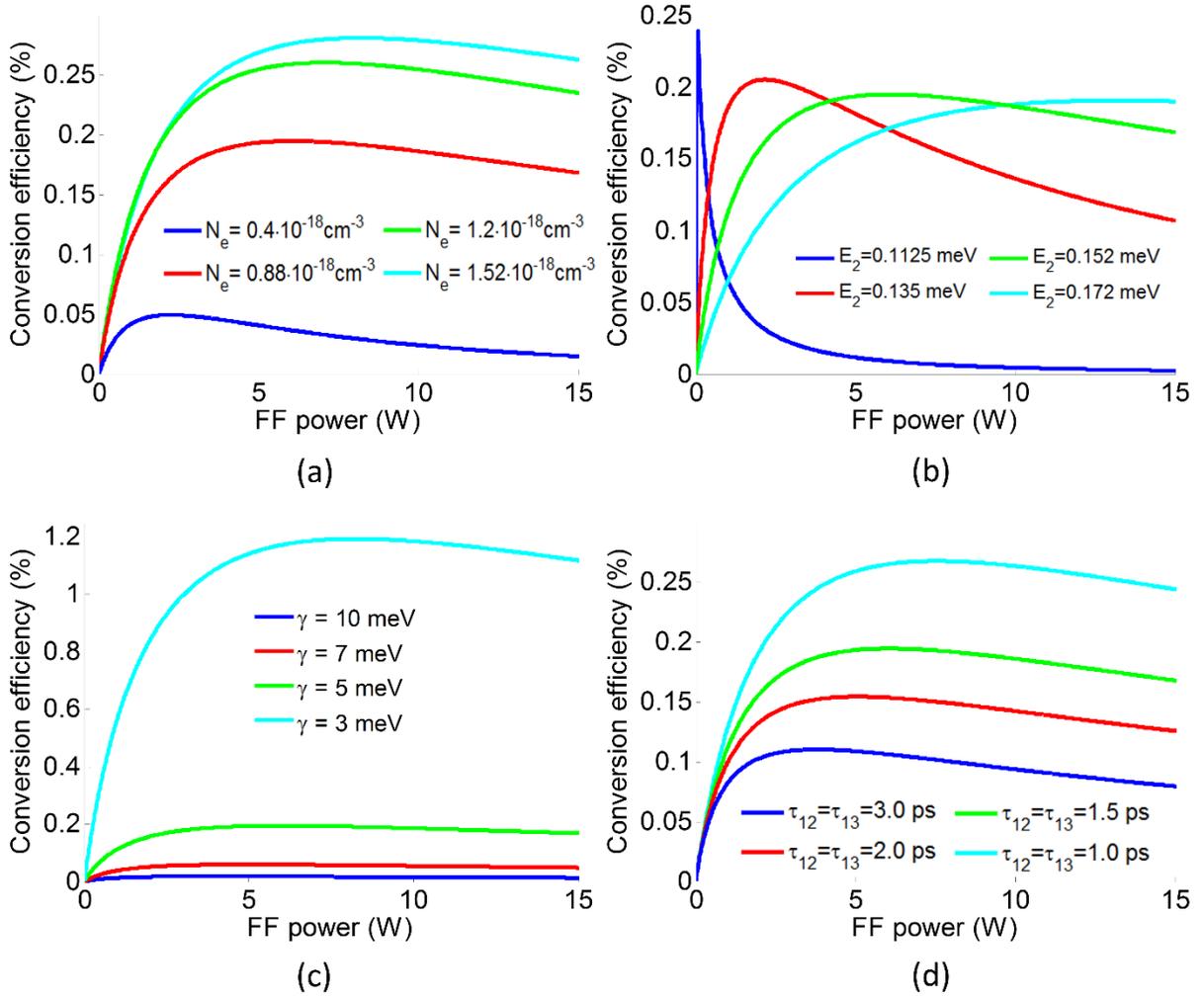

Figure 6 – Power conversion efficiency of the nonlinear metasurface shown in Fig. 1 versus the features of the MQW. (a) Doping level. (b) Energy of the second level. (c) Linewidth. (d) Relaxation time between the second and first energy bands. The MQW configuration employed in the simulations is described in the text. The doping level is set to $N_e = 8.8 \cdot 10^{-17}\,\mathrm{cm}^{-3}$. The structures are excited by an incident Gaussian beam with 17um of radius.